\documentclass[twocolumn,aps,prl,superscriptaddress,nofootinbib]{revtex4-1}
\usepackage{latexsym,amssymb,amsmath,epsfig,bm,psfrag}
\usepackage{color}
\usepackage[bookmarksnumbered,bookmarksopen,colorlinks,citecolor=red,linkcolor=blue]{hyperref}
\usepackage{graphicx}
\usepackage{times}
\usepackage{ulem,slashed}
\usepackage{soul}
\usepackage{subfiles} 

\hypersetup{
    colorlinks=true,
    linkcolor=blue,
    filecolor=magenta,
    urlcolor=cyan,
    pdftitle={Sharelatex Example},
    bookmarks=true,
}

\newcommand\E{\mathcal  E}

\newcommand\ve{\varepsilon}
\newcommand\Pe{\mathbb P}

\newcommand{\Fig}[1]{Fig.~\ref{#1}}
\newcommand{\Eq}[1]{Eq.~(\ref{#1})}

\newcommand\be{\begin{equation}}
\newcommand\ee{\end{equation}}

\newcommand{\trento}{{T$_{\rm R}$ENTo }}

\begin{document}

\title{On the collective flow puzzle in ultra-central heavy-ion collisions}

\author{Chen-Xi Liang}
\affiliation{Institute of Modern Physics, Fudan University, Shanghai 200433, China}

\author{Li Yan}
\email{cliyan@fudan.edu.cn}
\affiliation{Institute of Modern Physics, Fudan University, Shanghai 200433, China}
\affiliation{Key Laboratory of Nuclear Physics and Ion-beam Application (MOE), Fudan University, Shanghai 200433, China}
\affiliation{Shanghai Research Center for Theoretical Nuclear Physics, National Natural Science Foundation of China and Fudan University, Shanghai 200438, China}

\begin{abstract}
Using UCC-DiffHIC, an AI-accelerated event-by-event hydrodynamic model validated against hybrid simulations, we study flow in ultracentral Pb-Pb collisions. For suitable initial conditions, the calculated $v_2\{2\}/v_3\{2\}$ approaches unity in the most central collisions, consistent with experiment. An exact response decomposition shows that residual fluctuations suppress the ratio through $\Pe_3/\Pe_2<1$, where $\Pe_2$ and $\Pe_3$ are Pearson response correlations for elliptic and triangular flow, respectively. This resolves the apparent flow hierarchy without requiring a breakdown of linear response or additional effects such as nuclear deformation.
\end{abstract}
\maketitle

Relativistic heavy-ion collisions provide a unique laboratory for exploring the collective dynamics 
of strongly interacting matter under extreme conditions. A striking discovery at the Relativistic Heavy Ion Collider (RHIC) and the Large Hadron Collider (LHC) is the
{\it perfect}
fluidity of the produced quark-gluon plasma (QGP)~\cite{Heinz:2013th}, revealed by
the harmonic flow coefficients $v_n$~\cite{PhysRevD.46.229,AlverRoland2010,TeaneyYan2011,Gardim2012}
that characterize the azimuthal emission of 
particles.
These coefficients have been successfully described by hydrodynamic modeling with extremely small dissipations.

The success of relativistic viscous hydrodynamics has led to the flow paradigm~\cite{TeaneyYan2011,QiuHeinz2011,Gardim2012,Yan:2017ivm}, in which $v_n$ arise from the conversion of initial-state spatial anisotropies $\varepsilon_n$ through the collective expansion of the QGP fluid. In particular, for the lowest harmonics, $n=2$ and $3$, corresponding to elliptic and triangular flow respectively,
event-by-event hydrodynamic simulations establish an approximately linear response~\cite{Niemi:2012aj,Noronha-Hostler:2015dbi},
\begin{align}
\label{eq:lresp0}
V_n = \kappa_n \E_n + \delta_n\,,\quad n=2,3
\end{align}
where $V_n=v_n e^{in\Psi_n}$ and $\E_n=\ve_n e^{in\Phi_n}$ are the complex flow and eccentricity vectors, the response coefficient $\kappa_n$ is a real constant within a specified centrality class.
The residual fluctuations $\delta_n$ collect event-by-event fluctuations of 
the response that are not aligned with $\E_n$, including, for instance, nonlinear response contributions~\cite{TeaneyYan2012Nonlinear}.

Although this paradigm has achieved remarkable quantitative success across a broad range of collision energies, systems, and centralities, it is persistently challenged by the collective flow puzzle in ultra-central heavy-ion collisions (UCCs). UCCs are collisions with nearly vanishing impact parameter, so that the average nuclear overlap approaches azimuthal symmetry and the initial-state anisotropies are dominated by event-by-event fluctuations. While UCCs offer a particularly clean opportunity to probe initial-state fluctuations and nuclear geometry~\cite{Giacalone2020ObserveDeformation,Jia2022ShapeNuclei,Jia2022ShapeNucleiII},
measurements of two-particle correlations in ultra-central Pb-Pb collisions, down to the most central one in ten thousand events, have revealed an unexpected degeneracy between elliptic and triangular flow~\cite{CMS2014UltraCentral,CMS2014HigherHarmonics,ALICE2016AnisotropicFlow502,ATLAS2018UCC},
$
    v_2\{2\} \equiv \langle v_2^2\rangle^{1/2}
    \simeq v_3\{2\}\equiv \langle v_3^2\rangle^{1/2},
$
where $\langle\ldots\rangle$ denotes the average over collision events.
In contrast, the flow paradigm generally expects a hierarchy 
following from
the linear flow response relation. On the one hand, the hydrodynamic response to initial-state anisotropies is damped by viscous effects, with stronger damping for shorter-wavelength structures~\cite{TeaneyYan2012Nonlinear,Staig:2010pn}, generically yielding $\kappa_2 > \kappa_3$. On the other hand, a
purely fluctuating nuclear geometry typically has comparable initial-state anisotropies in UCCs, $\varepsilon_2 \simeq \varepsilon_3 \propto 1/\sqrt{N}$~\cite{Bhalerao:2011bp,Yan:2015fva}, with $N$ the number of independent sources contributing to the fluctuations, so one naturally has the form of a linear scaling,
\begin{align}
\label{eq:lresp1}
    \frac{v_2\{2\}}{v_3\{2\}} \simeq
    \frac{\kappa_2 \ve_2\{2\}}{\kappa_3\ve_3\{2\}} > 1\,.
\end{align}

The UCC flow puzzle has stimulated extensive investigations of possible missing physics in both the initial state and the subsequent dynamical evolution.
Proposed mechanisms include subnucleonic fluctuations ~\cite{Wang2026InitialStateStructures}, nuclear structure and deformation effects~\cite{Carzon2020,Xu2025OctupoleFluctuations}, nucleon correlations, modified pre-equilibrium dynamics, temperature-dependent transport coefficients,
stochastic hydrodynamic fluctuations~\cite{Kuroki2023HydroFluctuationsUCC}, and Bayesian-calibrated multistage models~\cite{Trajectum2021,Trajectum2022,ExTrEMe2023}. While these effects can modify the relative magnitudes of $v_2$ and $v_3$, a quantitative and unified resolution of the puzzle remains elusive. In particular, state-of-the-art multistage calculations retain a splitting $v_2\{2\}/v_3\{2\}\approx1.2$--$1.3$ in the ultra-central limit~\cite{Trajectum2022,ExTrEMe2023}.

Partly, the persistence of the puzzle stems from a computational barrier. Ultra-central classes occupy only a fraction of a percent of the inelastic nucleus-nucleus cross section. 
Accordingly, theoretical calculations of UCC flow require explicit 
event-by-event hydrodynamic and hadronic simulations at statistics that are prohibitive for conventional approaches. Most existing studies have therefore been forced to rely on approximations, especially the extrapolation of the linear scaling, 
\Eq{eq:lresp1}, into a regime where its validity to a large extent has 
remained untested so far. 
Recent advances in artificial intelligence, however, have enabled more efficient numerical approaches to hydrodynamic modeling, substantially reducing this computational barrier.

In this Letter, 
with an AI-based hydrodynamic modeling 
DiffHIC~\cite{Sun:2024lgo} 
improved by a UCC-specialized training 
stage, UCC-DiffHIC, we perform effective event-by-event hydrodynamic simulations of the most 
central Pb-Pb collisions at $\sqrt{s_{NN}}=2.76$~TeV 
and confront the corresponding experimental measurements
~\cite{CMS2014UltraCentral,CMS2014HigherHarmonics}. 
We notice that 
the ultra-central flow hierarchy presents 
a weak dependence on the beam energy~\cite{ATLAS2018UCC,ALICE2016AnisotropicFlow502,ALICE2018vn4}.
Our results exhibit a consistent decrease of the ratio $v_2\{2\}/v_3\{2\}$ towards unity as the
centrality decreases from $2.5$-$5\%$ to $0$-$0.02\%$, in agreement with data.
Decomposing the simulated response exactly as in \Eq{eq:lresp0} from event to event, 
we find that the flow ratio obeys the exact identity
\begin{align}
\label{eq:main}
\frac{v_2\{2\}}{v_3\{2\}}=\frac{\kappa_2\ve_2\{2\}}{\kappa_3\ve_3\{2\}} \times \frac{\Pe_3}{\Pe_2}
\end{align}
where $\Pe_2$ and $\Pe_3$ are the Pearson correlation coefficients. 
Because of the generic ordering $\Pe_3<\Pe_2$ 
due to a fluctuating source and dissipative damping, finite residual fluctuations inevitably reduce the ratio below 
its linear-scaling estimate. 
The reduction could be sufficiently 
strong regarding fluctuating initial
state in UCCs, and
eventually compensates for the linear response and results in flow degeneracy in UCCs.
The reduction mechanism is not a consequence of 
specific initial-condition parameters or 
transport coefficients. 


The residual fluctuations in \Eq{eq:lresp0}, although unavoidably present in any event-by-event simulations, have, to our knowledge, never been analyzed 
in the context of UCC flow. More importantly, their role becomes apparent only through event-by-event simulations of these extremely rare collisions. Such studies were previously computationally impractical, 
but are now feasible through realistic, high-statistics simulations enabled by AI-based hydrodynamic modeling, underscoring the growing significance of precision studies in high-energy heavy-ion collisions.

{\it AI-based hydrodynamic modeling} -- 
The reference calculation consists of a conventional event-by-event hybrid chain. For each Pb-Pb collision at $\sqrt{s_{NN}}=2.76$~TeV, a fluctuating \trento entropy-density profile $s(x,y)$~\cite{Moreland:2014oya} 
is evolved by the viscous hydrodynamic code MUSIC~\cite{Schenke:2010rr,Schenke:2012wb}, 
with a lattice-QCD equation of state matched to a hadron resonance gas~\cite{Bazavov:2014HotQCD}, a constant specific shear viscosity $\eta/s=0.1$, and a peaked bulk-viscosity profile.
Hadrons are sampled on the switching hypersurface by Monte Carlo particlization and rescattered in the UrQMD hadronic afterburner~\cite{Bass:1998ca,Bleicher:1999xi}. 
Eventually, the hybrid modeling converts initial entropy-density profile, $s(x,y)$, to 
$\rho(p_T,\phi)$, which is the accepted charged-particle distribution of identified $\pi^\pm$, $K^\pm$, $p$, $\bar p$ binned on a $64\times64$ grid in azimuth and transverse momentum~\cite{note:accept}.

\begin{figure}
\begin{center}
    \includegraphics[width=0.45\textwidth]{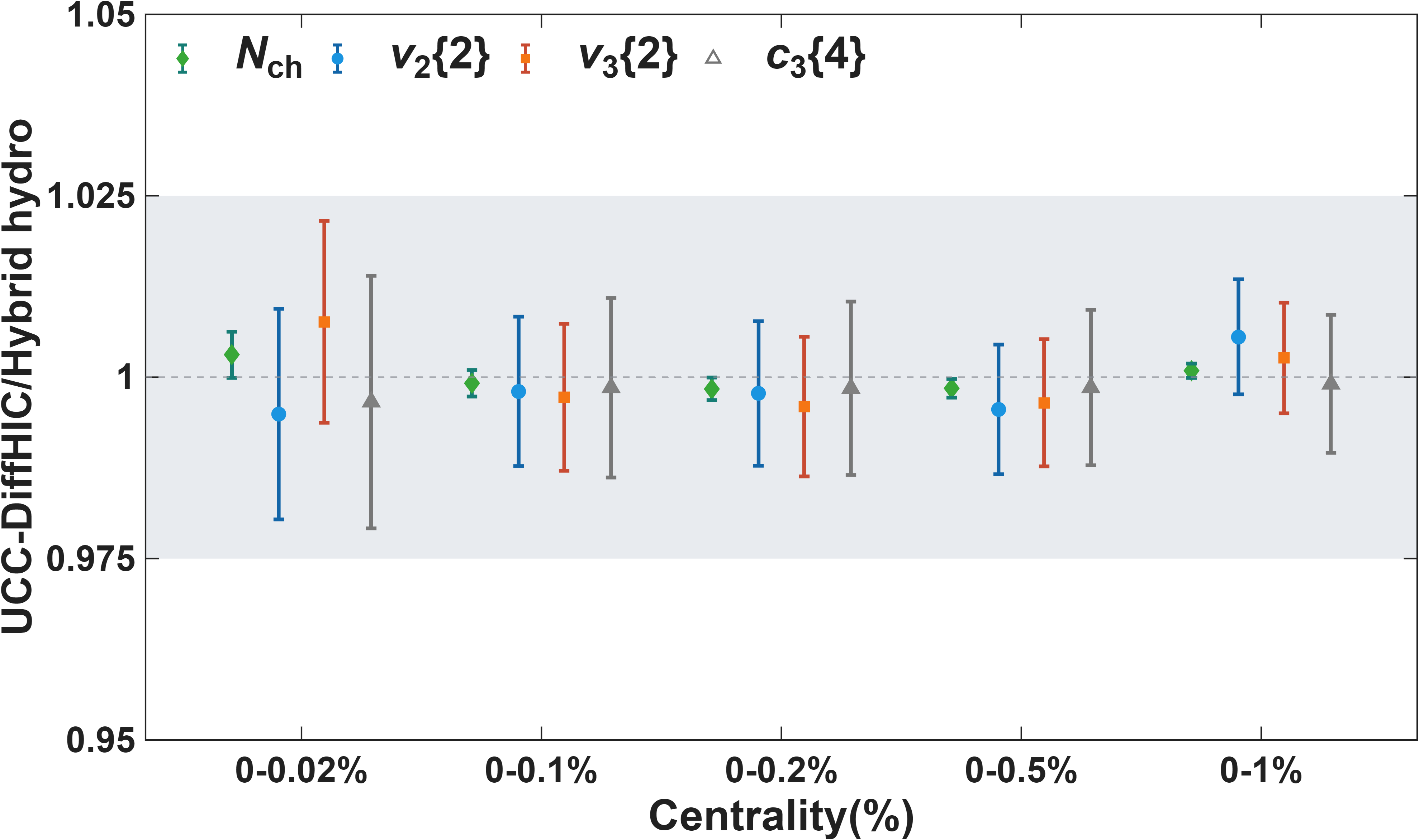}
    \caption{Validation of UCC-DiffHIC against the hybrid hydro chain in terms of 
    the charged particle multiplicity, $v_2\{2\}$, $v_3\{2\}$ and 
    $c_3\{4\}$ in UCCs. With respect to the ratios of these observables from DiffHIC and the hybrid hydro chain, 
    the characterization of flow in UCCs are reliable within $2.5\%$ (the grey band).  
    }
    \label{fig:validate}
\end{center}
\end{figure}

DiffHIC~\cite{Sun:2024lgo} employs a conditional score-based diffusion model~\cite{Ho:2020DDPM,Song:2021SDE} to learn directly the mapping implemented by the full hybrid chain,
\be
\label{eq:cascade}
s(x,y)\ \xrightarrow{\ \mathrm{DiffHIC}\ }\ \rho_{\rm base}(p_T,\phi)\,, 
\ee
achieving high 
fidelity with substantially improved computational efficiency. More specifically, a U-Net with self-attention predicts the noise field of a variance-preserving stochastic differential equation, conditioned on the initial-condition image and global transport labels. The sampling proceeds through predictor--corrector integration of the reverse process. DiffHIC reproduces the spectra and anisotropic flow generated by the hybrid chain across its training ensemble.
Generating a full spectrum for a typical event takes $\sim\!0.1$s on a consumer-grade GPU, compared with ${\cal O}(10^2)$ single-core CPU-minutes for the full hybrid hydrodynamic 
calculation. This corresponds to a speedup of order $10^5$~\cite{Sun:2024lgo}, making statistically meaningful UCC samples computationally feasible. The released DiffHIC weights are kept frozen throughout this work.

The baseline DiffHIC yields flow predictions with reported
uncertainties $\lesssim5\%$ over a broad centrality range, its
$\sim10^4$ minimum-bias training events, however, 
under-samples the UCCs with 
centrality below $1\%$.
In particular,
a reliable description of UCC flow requires control of event-by-event flow fluctuations and initial--final correlations, beyond agreement with the average particle spectrum. We therefore introduce UCC-DiffHIC, an improved model for the UCC regime. It augments the baseline with a geometry-conditioned residual network trained exclusively on ultra-central events. Acting on the baseline spectrum, the network predicts
\be
\label{eq:corrector}
z_{\rm corr} = z_{\rm base} + g(c)\, r_{\rm max}\tanh f_\theta\!\left(z_{\rm base}, q\right),
\ee
where $f_\theta$ is a compact two-level residual U-Net with $8.8\times10^5$ parameters, $r_{\rm max}=0.35$ bounds the correction, and $q$ is an 18-component vector of initial-geometry summaries (centrality, impact parameter, entropy, size, and the complex eccentricity vectors, $\E_2,\dots,\E_5$) injected through feature-wise modulation. When applied at inference,
the hard gate $g(c)=\mathbf{1}_{[0,1\%]}(c)$ leaves the baseline
output unchanged outside the UCC domain. The network is trained in a supervised manner on approximately $1.6\times10^4$ {\it initial-final} events within $1\%$ centrality,
 for which the same initial condition is processed by both the baseline and the full hybrid chain. 
 Consequently, the rarest centrality bins, such as $0$-$0.02\%$, 
are enriched and reweighted. The training objective ${\cal L}$ combines terms for the pixel-wise spectrum, the multiplicity, the magnitudes $v_{2},\dots,v_5$, and the complex vectors, $V_2,\ldots, V_5$.
Besides, to improve the description of event-by-event fluctuations, we apply a trainable residual correction to the baseline output spectrum. The fluctuation structure of the corrected predictions is assessed through the fourth-order flow cumulants $c_n\{4\}$, which provide constraints beyond the second moments. Such restrictions are implemented by optimizing ${\cal L}_{\rm response}$ in ${\cal L}$ . The agreement with the reference values within the uncertainties quoted supports the consistency of the corrected model with the higher-order fluctuations probed by these observables. So, the correction is constrained by precisely the event-by-event information that defines the UCC puzzle.

The model is validated with approximately 5000 
collision events with centrality below $1\%$.
A standard set of 
initial-state parameterization of \trento model (Set I, to be specified below) is taken for primary training and validation, with analogous procedures 
applied to other sets of parameters.
We examine the charged-particle multiplicity $N_{\rm ch}$, $v_2\{2\}$, $v_3\{2\}$, and $c_3\{4\}=\langle v_3^4\rangle-2\langle v_3^2\rangle^2$,
which probe the average multiplicity, two- 
and four-particle correlations, respectively. 
As shown in \Fig{fig:validate}, the central values agree with the 
hybrid chain within 2.5\%, with the uncertainty increases for higher order correlations.
These errors will be propagated and treated 
as systematic uncertainties in the following analyses.
Further details of the UCC-DiffHIC are given in the Supplemental Material.

\begin{figure}[t]
\begin{center}
    \includegraphics[width=0.45\textwidth]{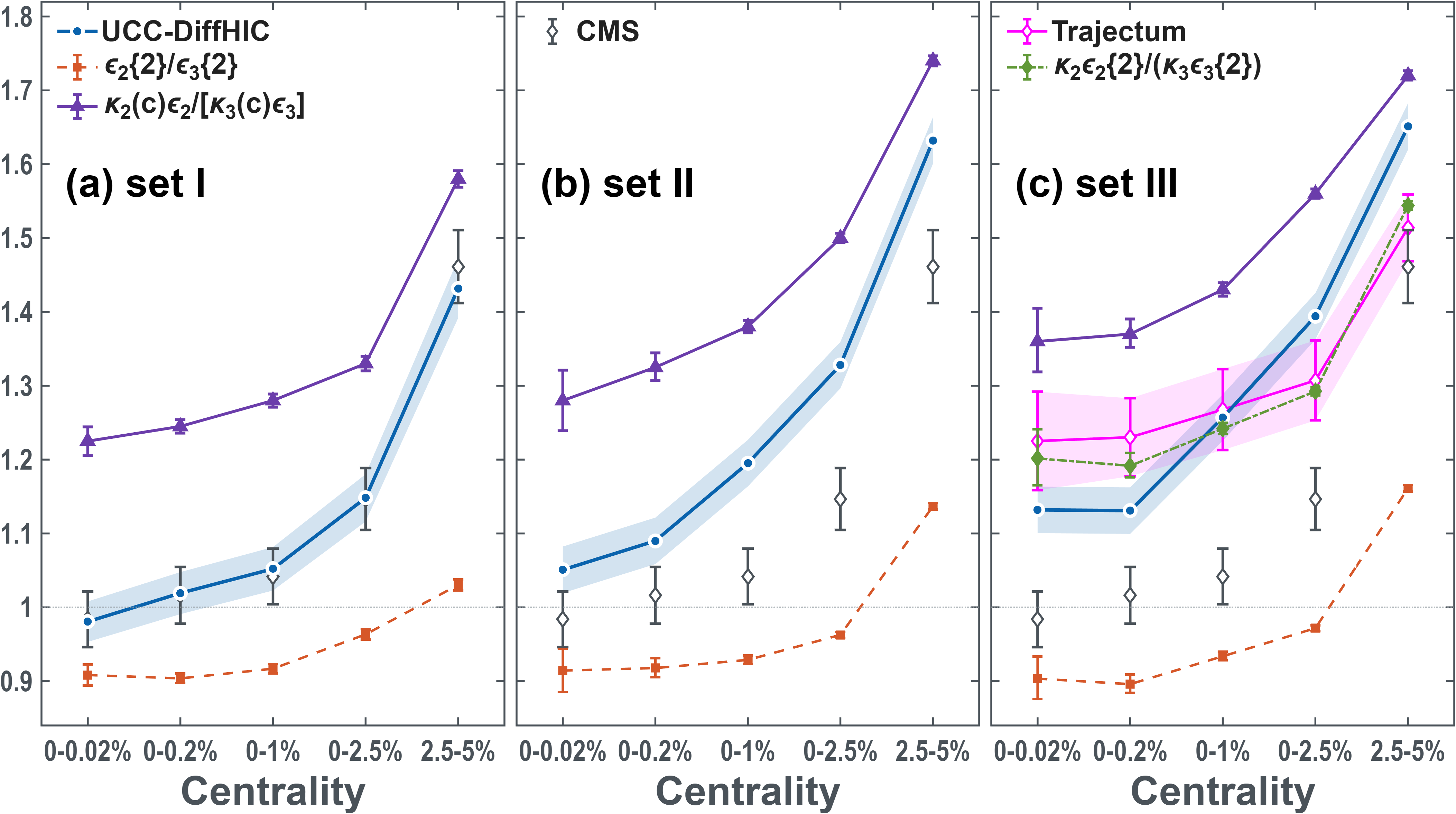}
    \caption{ 
    Flow ratio $v_2\{2\}/v_3\{2\}$ from UCC-DiffHIC for three sets of initial-state parameterizations. The event-by-event results (blue) are compared with the initial-eccentricity ratios (red), the linear-scaling estimates (purple), and CMS measurements (black open symbols). The suppression relative to linear scaling arises from $\Pe_3/\Pe_2<1$ and becomes stronger in more central collisions. Set I closely follows the CMS data, whereas Set III retains a stronger splitting that is consistent with previous Trajectum calculations (magenta).
    A linear scaling of initial eccentricity ratios with one averaged value of $\kappa_3/\kappa_2$ in the central collisions in shown as the green line in
    panel (c).
    }
    \label{fig:vnratio}
\end{center}
\end{figure}

{\it Flow ratio from UCC-DiffHIC.} --
Using the large event samples generated with UCC-DiffHIC, 
we calculate the ratio $v_2\{2\}/v_3\{2\}$ 
from event-by-event simulations, with the results shown in \Fig{fig:vnratio}. 
In most experiments, such as CMS, the centrality class of nucleus-nucleus collisions is determined 
via the probability distribution of charged particle production. In our simulations, considering 
the approximately linear correlation between initial-state entropy and charged particle multiplicity, $N_{\rm ch}$,
we select centrality by initial-state entropy. For instance, the most central collisions correspond
to initial states with largest entropy. 
We have checked that, 
classifying centrality by initial-state entropy,
or $N_{\rm ch}$, has only a marginal effect on the flow ratio $v_2\{2\}/v_3\{2\}$. 
In these simulations, the fluid-dynamical expansion is held fixed, in terms of 
the transport coefficients and equation of state, etc., 
while three different sets of initial-state parameters for the \trento model are considered for comparison.
Set I [panel (a)] corresponds to the standard \trento parameter choice used 
in the original DiffHIC training~\cite{Sun:2024lgo}. Set II [panel (b)] adopts the Bayesian parameter estimates obtained with Trajectum but omits the initial free-streaming stage~\cite{Trajectum2022}. 
Set III [panel (c)] uses the same parameters as in Ref.~\cite{Trajectum2022}, namely, same \trento 
parameters as in Set II but including the initial free-streaming stage. 
Further details of the \trento parameters are given in the Supplemental Material.

Notably, the ratio $v_2\{2\}/v_3\{2\}$ obtained with Set I agrees remarkably well with the CMS measurement, 
consistently approaching unity in the most central collisions. 
In contrast, simulations employing the Bayesian parameter sets, 
either with or without initial free streaming, 
fail to reproduce the UCC flow degeneracy. We note that
these Bayesian posterior parameters were 
calibrated with respect to a variety of 
flow observables in 
centrality classes 
excluding the dedicated UCC measurements, as noted previously~\cite{Trajectum2022,ExTrEMe2023}.
Although a direct quantitative comparison is not expected, our Set III result is compatible with previous Trajectum calculations for Pb-Pb collisions at $\sqrt{s_{\rm NN}}=5.02$~TeV with a temperature-dependent $\eta/s$. 

For all three parameter sets, we also show the linear-scaling estimate, 
$\kappa_2\ve_2\{2\}/\kappa_3\ve_3\{2\}$, as the purple solid lines in \Fig{fig:vnratio}. 
While the ratio of initial eccentricities,
$\ve_2\{2\}/\ve_3\{2\}$, of the three sets are comparable, 
the linear scaling differs and always leads to larger flow ratio 
$v_2\{2\}/v_3\{2\}$ than those from event-by-event simulations. More importantly, 
unlike the event-by-event simulations, 
this simple linear-scaling approximation yields a clear hierarchy, $v_2\{2\}>v_3\{2\}$, which persists 
across all three parameter sets in UCCs. 
It is interesting that the ratio $v_2\{2\}/v_3\{2\}$ obtained 
with Trajectum 
approximately follows 
the linear-scaling estimate based on the ratio of initial eccentricities, 
using a constant, averaged value of $\kappa_2/\kappa_3$ in UCCs.

The apparent discrepancy between the UCC flow ratio $v_2\{2\}/v_3\{2\}$ obtained from event-by-event simulations and the linear-scaling estimate $\kappa_2\ve_2\{2\}/\kappa_3\ve_3\{2\}$ does not indicate a breakdown of the flow paradigm or, more specifically, of \Eq{eq:lresp0}. Instead, it reflects the imperfect event-by-event correlation between the final-state flow and the initial-state geometry. Defining $\kappa_2={\rm Re}\langle V_2\E_2^*\rangle/\ve_2\{2\}$ and $\kappa_3={\rm Re}\langle V_3\E_3^*\rangle/\ve_3\{2\}$, \Eq{eq:lresp0} implies that the residual fluctuations $\delta_n$ contribute naturally to the two-particle correlations according to
\begin{align}
\langle v_n^2\rangle = \kappa_n^2 \langle \ve_n^2\rangle + \langle |\delta_n|^2\rangle \,,\qquad n=2,3\,.
\end{align}
The effect of the residual fluctuations can be equivalently characterized by the Pearson correlation coefficients,
\begin{align}
\Pe_n \equiv \frac{{\rm Re}\langle V_n\E_n^*\rangle}{\langle v_n^2\rangle^{1/2}\langle\ve_n^2\rangle^{1/2}}
= \left(1-\frac{\langle|\delta_n|^2\rangle}{\langle v_n^2\rangle}\right)^{1/2}\,,\qquad n=2,3\,.
\end{align}
Thus, finite residual fluctuations imply $\Pe_n<1$ unless the linear correlation is perfect. Moreover, the event-by-event flow ratio differs from the eccentricity-based estimate by a factor $\Pe_3/\Pe_2$,
as shown in \Eq{eq:main}.
The comparisons in \Fig{fig:vnratio} demonstrate the importance of this correction. For all three initial-state parameter sets, the eccentricity-based linear-scaling estimate alone yields in UCCs a clear hierarchy, $v_2\{2\}>v_3\{2\}$, 
which is 
systematically reduced, 
with the dominant effect arising from the imperfect Pearson correlations through $\Pe_3/\Pe_2<1$. The initial conditions also influence the response coefficients and hence the eccentricity-based contribution, but this constitutes a secondary effect. For Set I, the reduction associated with $\Pe_3/\Pe_2$ compensates for the linear-scaling enhancement, bringing the flow ratio into close agreement with the CMS measurement and toward unity in the most central collisions. Sets II and III exhibit a less complete compensation and therefore retain larger flow ratios. 

\begin{figure}[t]
\begin{center}
    \includegraphics[width=0.45\textwidth]{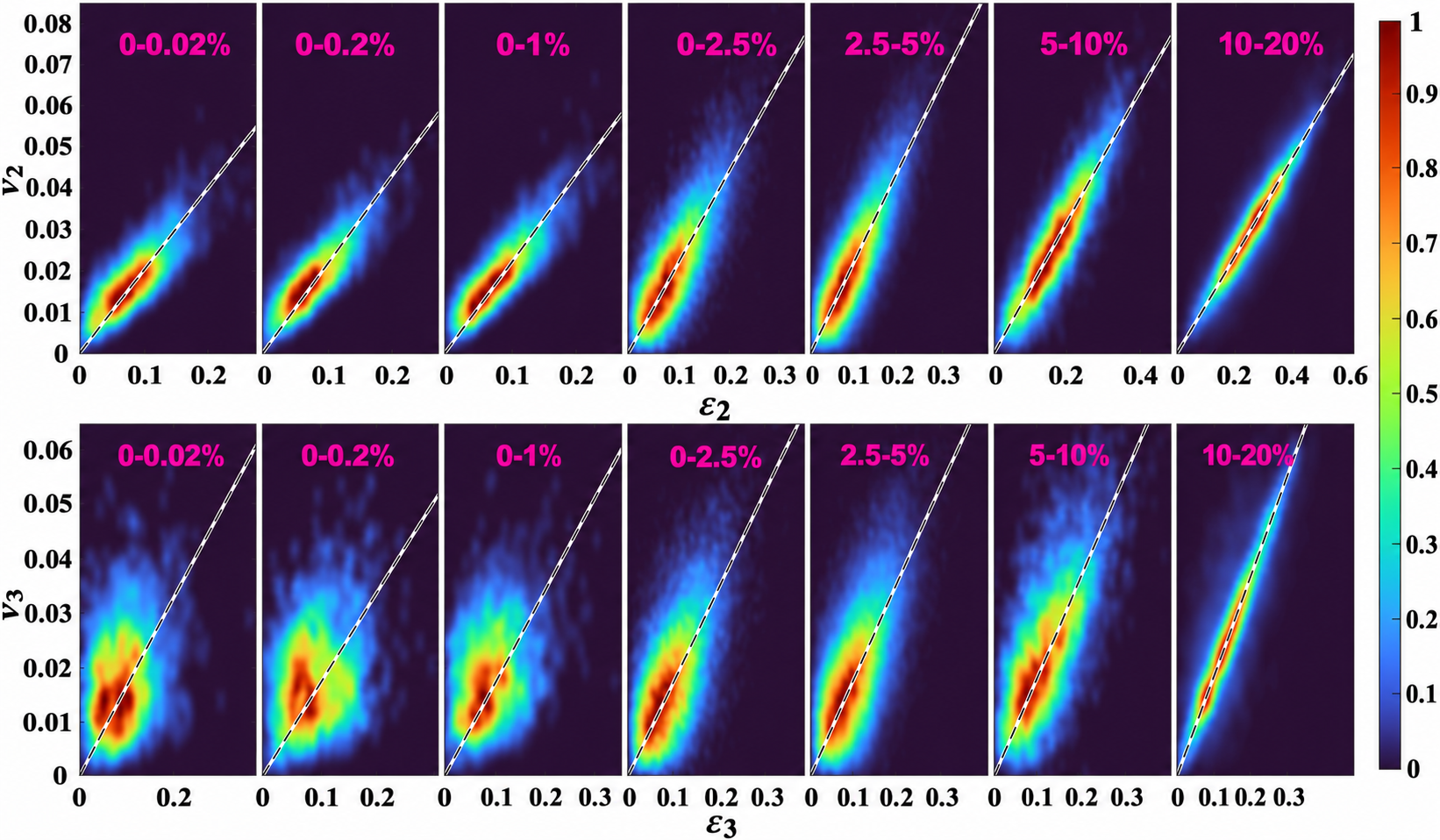}
    \caption{ Event-by-event correlation between $v_2$ and $\ve_2$ (upper panel),
    and between $v_3$ and $\ve_3$ (lower panel),
    for various centralities. The linear correlation between flow and initial eccentricity gets weaker
    consistently towards ultra-central 
    collisions, for both elliptic flow and the triangular flow. 
    Colors denote the normalized event density.
    }
    \label{fig:vnen}
\end{center}
\end{figure}

This interpretation is directly supported by the event-by-event distributions for Set I, shown in \Fig{fig:vnen}. In the most central bins, both $v_2$ and $v_3$ exhibit substantial event-by-event variations at fixed $\ve_2$ and $\ve_3$, respectively, demonstrating that the initial eccentricity alone does not fully determine the final-state flow. The dispersion around the linear-response trend is particularly pronounced for $v_3$ versus $\ve_3$, indicating a weaker linear correlation and hence a smaller $\Pe_3$ relative to $\Pe_2$. Toward less central collisions, both distributions become progressively narrower and more closely aligned with the linear-response trends. For elliptic flow, this tightening is further assisted by the emergence of a dominant average elliptic geometry, whereas triangular flow remains more strongly affected by fluctuations~\cite{Niemi:2012aj,Noronha-Hostler:2015dbi}. 
Similar behavior is found for Sets II and III, with quantitative differences in the strength of the flow--eccentricity correlations (see the Supplemental Material). 

Furthermore, a parametric argument suggests that the inequality $\Pe_3/\Pe_2<1$ arises naturally from the fluctuating initial geometry and viscous damping, 
in consistent with the event-by-event hydrodynamic simulations in Fig.~\ref{fig:vnen}. We first isolate the dominant nonlinear contributions to the residual fluctuations in \Eq{eq:lresp0} by writing, for $n=2$ and $3$, respectively, 
\begin{align}
V_2 =& \kappa_2 \E_2 + \kappa_{2,11} \E_{1,3}\E_{1,3} + \delta_2'\,,\nonumber\\
V_3 =& \kappa_3 \E_3 + \kappa_{3,12} \E_{1,3}\E_{2} + \delta_3'\,.
\end{align}
We have checked that the inclusion of nonlinear contributions substantially improves correlations between flow and initial state, for instance, $V_3$ and
$\kappa_3 \E_3 + \kappa_{3,12}\E_1\E_2$ are apparently linear correlated with a Pearson correlation around 0.7 in the most central collisions, comparing to roughly 0.4 for $V_3$ and $\kappa_3 \E_3$.
We implicitly adopt the cumulant definition of the initial-state eccentricities~\cite{TeaneyYan2011}, so that the terms in each response relation are statistically uncorrelated 
in the ultra-central limit. The dipole asymmetry $\E_{1,3}$, like $\E_3$, arises purely from geometric fluctuations. In contrast, $\E_2$ receives contributions from both geometric fluctuations and the average geometry, except in the ultra-central limit.
For pure geometrical fluctuations, there exist crude scalings $\langle\ve_2^2\rangle\sim\bar{\ve}_2^2+1/N$ and $\langle\ve_1^2\rangle=\langle\ve_3^2\rangle\sim1/N$. Together with the viscous-damping ordering 
of the response coefficients~\cite{Staig:2010pn,TeaneyYan2012Nonlinear},
$\kappa_{n,m}/\kappa_{\rm ideal}\sim e^{ - \left(\frac{n+m}{2}\right)^2 \Gamma}\,,$
where $\kappa_{\rm ideal}$ is the response coefficient in the ideal-fluid limit of an infinitely large system, $\Gamma>0$ characterizes the dissipative effects, and $n$ and $m$ are integers associated with the cumulant definitions, these scalings yield the following analytic estimate in the ultra-central limit:
\begin{align}
\frac{\Pe_3}{\Pe_2}\sim
\sqrt{\frac{1+e^{-8\Gamma}/N}{1+e^{2\Gamma}/N}}<1\,.
\end{align}
For a finite but small $\bar{\ve}_2$, this estimate is modified, but the inequality remains valid. Further details are given in the Supplemental Material.

The precise value of $\Pe_3/\Pe_2$ depends on the detailed fluctuating geometry, as determined by the initial-state model parameters, and on the fluid expansion through the Knudsen number and system size. The conditions realized in UCCs can therefore yield a value of $\Pe_3/\Pe_2$ that compensates for the linear scaling of the initial eccentricities and drives the flow ratio toward unity, as observed for Set I in the present study. Nevertheless, it is important to emphasize that $\Pe_3/\Pe_2<1$ is generic, even without additional physical inputs such as nuclear deformation.

{\it Summary and discussion.} --
We have presented UCC-DiffHIC, an AI-based event-by-event hydrodynamic framework adapted to the fluctuation-dominated UCC regime through targeted training. It reproduces the reference hybrid-chain calculations for the charged-particle multiplicity and two- and four-particle flow observables with central-value deviations within $2.5\%$, enabling statistically meaningful analyses of exceptionally rare collision classes.

As a direct application of UCC-DiffHIC, we carried out event-by-event 
analyses of the flow ratio. 
For three sets of initial-state parameterizations, residual response fluctuations systematically reduce the event-by-event ratio $v_2\{2\}/v_3\{2\}$ relative to the linear-scaling estimate through $\Pe_3/\Pe_2$. The reduction gets stronger toward the most central collisions and varies quantitatively with the initial-state 
geometrical fluctuations.
For Set I, this reduction compensates for the linear-scaling 
enhancement and reproduces the experimentally observed flow degeneracy. 
Application of UCC-DiffHIC to 
higher-order UCC flow cumulant 
also help to constrain nuclear-structure interpretations. 
While a sufficiently strong 
octupole deformation 
could enhance 
$v_3\{4\}/v_3\{2\}$ above the fluctuation limit, 
our UCC-DiffHIC simulations gives roughly a constant $v_3\{4\}/v_3\{2\}$ compatible with $\sqrt{4/\pi-1}$ from 5\% to UCCs
(see details in Supplemental Material),
in consistent with the 
recently reported ALICE Run~3 result~\cite{Likmeta:2026clq}.

Our results therefore demonstrate that, 
the observed flow degeneracy in UCCs requires neither 
a breakdown of the flow-response paradigm nor 
a sizable static 
nuclear deformation as the necessary origin, 
although smaller or fluctuating contributions cannot be excluded. Our analysis thus accounts for the apparent UCC discrepancy within the established event-by-event framework. The puzzle arises when residual response fluctuations are under-estimated from the mean linear-scaling relation.
Beyond this specific physics problem of flow puzzle, 
UCC-DiffHIC demonstrates how AI-based 
modeling can influence the precision studies 
of QGP dynamics and nuclear structure.

{\it Acknowledgements.} --
This study is supported in part by the National Natural Science Fundation of China (NSFC) through Grant No. 12375133 and 12347106.

\bibliography{references}

\end{document}